\documentclass[5p,twocolumn,final]{elsarticle}

\usepackage{amssymb,amsmath,amsfonts}
\usepackage{float,graphicx,ulem}
\newcommand{\beq}{
\begin{equation}
}
\newcommand{\eeq}{
\end{equation}
}

\begin{document}

\begin{frontmatter}
\title{Particle-based modelling of aggregation and fragmentation processes: Fractal-like aggregates}
\author{Jens C. Zahnow\corref{cor1}} 
\author{Joeran Maerz} 
\author{Ulrike Feudel}
\cortext[cor1]{Corresponding author: zahnow@icbm.de}
\address{Theoretical Physics/Complex Systems, ICBM, University of Oldenburg, 26129 Oldenburg, Germany}
% ---------------------------------------------------------------------------------------------------------------
%				ABSTRACT
% ---------------------------------------------------------------------------------------------------------------
\begin{abstract}
The incorporation of particle inertia into the usual mean field theory for particle aggregation and fragmentation in fluid flows is still an unsolved problem. We therefore suggest an alternative approach that is based on the dynamics of individual inertial particles and apply this to study steady state particle size distributions in a 3-d synthetic turbulent flow. We show how a fractal-like structure, typical of aggregates in natural systems, can be incorporated in an approximate way into the aggregation and fragmentation model by introducing effective densities and radii. We apply this model to the special case of marine aggregates in coastal areas and investigate numerically the impact of three different modes of fragmentation: large-scale splitting, where fragments have similar sizes, erosion, where one of the fragments is much smaller than the other and uniform fragmentation, where all sizes of fragments occur with the same probability. We find that the steady state particle size distribution depends strongly on the mode of fragmentation. The resulting size distribution for large-scale fragmentation is exponential. As some aggregate distributions found in published measurements share this latter characteristic, this may indicate that large-scale fragmentation is the primary mode of fragmentation in these cases. \end{abstract}


\end{frontmatter}

% ---------------------------------------------------------------------------------------------------------------
%				INTRODUCTION
% ---------------------------------------------------------------------------------------------------------------
\section{Introduction}
\label{sec:Introduction}
In recent years there has been a great effort to investigate the advection of inertial particles in fluid flows \cite{Babiano2000,Nishikawa2001,Bec2003,bec2005,Vilela2007,Zahnow2008}. Understanding the behavior of inertial particles like aggregates, dust or bubbles moving in incompressible flows plays an important role in such diverse fields as cloud physics \cite{PruppacherKlett}, engineering \cite{Crowe}, marine snow and sediment dynamics \cite{Jackson1990,Mcanally2002} as well as wastewater treatment \cite{Zhang2003}. The dynamics of individual inertial particles is dissipative. This leads to a behavior that is very different from passive tracers, for example to a preferential accumulation in certain regions in space, i.e. on attractors \cite{Maxey1987,Benczik2003,Do2004}. Previous studies concentrated mainly on non-interacting particles, in spite of the fact that accumulation leads unavoidable to mutual interactions of different kinds. \\
It is well known that as a result of collisions between particles, aggregates can be formed that consist of a large number of primary particles. In many areas of science the formation of such aggregates and their break-up due to shear forces in the flow plays a very important role, e.g. in sedimentation of particles in oceans and lakes \cite{Winterwerp1998}, chemical engineering systems such as solid-liquid separation \cite{Spicer1996,Baebler2008}, aggregation of marine aggregates \cite{Thomas1999} and flocculation of cells \cite{Han2003}. \\
Most approaches of aggregation and fragmentation models are based on the pioneering work of Smoluchowski \cite{Smoluchowski1917} and use usually a mean field approach with kinetic rate equations to model these processes (see e.g.  Jackson \cite{Jackson1990}). However, for particles with inertia a field theory for the particle velocity has not yet been formulated. The existence of caustics, meaning that the dynamics of inertial particles  would lead at some points to a multi-valued particle velocity field \cite{Wilkinson2005,Derevyanko2007}, has prevented such an approach so far. While attempts have been made to incorporate particle inertia in approximate ways in a mean field approach \cite{Ayala2008_1,Ayala2008_2}, no completely satisfying solution has been found yet. \\
Here we therefore choose a different, individual particle based approach, where the dynamics of finite size particles are taken directly into account. The approach has recently been introduced in Ref. \cite{Zahnow2008_2}, and discussed in more detail with respect to different flows in \cite{Zahnow2008_3}. In the present study we adopt this approach to study the long-term behavior of particle size distributions that develop from a balance between aggregation and fragmentation. In particular, we examine the influence of fragmentation and aggregate structure on these size distributions. \\
In most previous works the particles were considered to be spheres with a specific density. In many realistic cases, for example for marine aggregates, this is a crude approximation. The complex structure of particles can have a great influence on particle dynamics as well as aggregation and fragmentation processes. Both the actual motion of aggregates and the probabilities for aggregation and fragmentation are influenced by the structure of the particles. \\
In the context of a mean field approach, a complex particle structure has been incorporated in the past in terms of a density modification for the particles, e.g. by Kranenburg \cite{Kranenburg1994} or Maggi et al. \cite{Maggi2007}. However, so far there are very few attempts to treat this problem for inertial particles in a flow. Wilkinson et al. \cite{Wilkinson2008} used a model for fractal particles in an aggregation model for dust particles during planet formation. Our present work expands the consideration of spherical particles to model more realistic aggregates. We focus specifically on the problem of aggregation and fragmentation in systems where the aggregates can  be described as having a \textit{fractal-like structure}, as is for example the case for marine aggregates \cite{Kranenburg1994}. By this we mean that on average there exists a power-law relationship between the characteristic length and the mass of such aggregates.  The exponent of the power-law is called the fractal dimension. Such a characterization in terms of a fractal dimension leads to a modification of the radii and effective densities of the aggregates compared to a solid sphere of the same mass.  Nevertheless, we still treat them as effectively spherical for the particle motion, allowing us to apply the Maxey-Riley equations of motion \cite{Maxey1983} with modified parameters. \\
In this work we choose a parameterization of our model for the case of a suspension of marine aggregates in the ocean. In this way we can study our modeling approach for a specific case, but we emphasize that our model is a general one that can in principle be used for a wide range of applications where aggregation and fragmentation of solid particles plays a role. The concept of a fractal-like structure has been found to be a reasonable first approximation in many different applications, ranging from colloidal systems to the flocculation of cells \cite{Logan}. A different application would require a different parametrization of the model, but the general approach introduced here would remain the same.  \\
Since the fractal dimension of marine aggregates can vary greatly in natural systems \cite{Maggi2007}, we examine its effect on the steady state particle size distributions in our model. We find that while the shape of the size distributions does not depend strongly on the fractal dimension, the average particle size and relaxation time towards the steady state depend strongly on this parameter.\\
Even though to a certain extent methods from dynamical systems theory can usefully be applied, we mention that the entire problem is much more complex than that of any usual dynamical system. While particles of a single size move on specific attractors, aggregation and fragmentation lead to repeated transitions from one attractor to another one, depending on the aggregate size. The skeleton of the new dynamics is therefore a superposition of the different attractors, with transient motion in between. The structure of the individual attractors and their superposition in turn influence the aggregation probabilities due to different local concentrations of particles. Fragmentation is also affected by the particle dynamics, because shear forces can be locally different in the flow. Therefore, break-up may depend on whether an attractor for a certain particle size lies in a region with high shear or not.\\
We show that the combination of aggregation and fragmentation of fractal-like aggregates, superimposed on inertial advection dynamics, leads to a convergence to a steady state in the particle size distribution. This steady state is unique for a given set of parameters. Mainly, we compare three different types of splitting, uniform fragmentation, erosion and large-scale fragmentation. These splitting modes differ in the size of the fragments that are created during break-up. While erosion leads to one large and one relatively small fragment, large-scale fragmentation leads to two fragments of similar size. We find that the transient dynamics as well as the size distribution in the steady state depend strongly on the splitting mode. The steady state size distribution found for large-scale fragmentation conforms best to observation reported in the literature for the break-up of marine aggregates in tidal areas  \cite{Lunau2006}, indicating that this may be the primary mode of fragmentation in these cases.\\
Section \ref{sec:Model} describes the complete model for advection, aggregation and fragmentation that is used in this work. The equations of motion for heavy spherical particles (Stokes equation) are used, but with modified parameters to take a fractal-like structure into account. Rules governing the aggregation and fragmentation are introduced. Finally, the model is applied to a simple 3-d synthetic turbulent flow field.\\
Section \ref{sec:Results} then presents a complete analysis of the influence of all major system parameters on the resulting steady state size distributions, the average aggregate size in steady state and the relaxation time towards the steady state. Namely, these parameters are aggregate strength, fractal dimension of the aggregates and total particle mass in the system. \\
Section \ref{sec:Conclusions} contains a discussion of the limitations of the model and the conclusions. 

% ---------------------------------------------------------------------------------------------------------------
%				MODEL
% ---------------------------------------------------------------------------------------------------------------

\section{Advection, Aggregation and Fragmentation Model}
\label{sec:Model}
In this section we will present the modeling approach used in this study, that describes the motion, aggregation and fragmentation of finite-size particles. Firstly, the equations of motion used for the advection of spherical particles heavier than the surrounding fluid are presented. Secondly, a model to account for the fractal-like structure of real aggregates is described. Thirdly, a full model to include aggregation and fragmentation in this context is introduced. Finally, a simple 3-d synthetic turbulent flow field is chosen, that will be used to study the aggregation and fragmentation model in detail.

% --------------------------- Equations of Motion----------------------------------------------------------------
\subsection{Equations of Motion for Spherical Particles}
\label{sec:Equations}
For simplicity, we consider all primary (smallest, unbreakable) particles to be spherical and denser than the ambient fluid. We emphasize that the equations of motion presented here for spherical particles will in the following also be used to describe the motion of aggregates which usually can not be assumed to be spherical \cite{Kranenburg1994}. However, to account for properties related to the fractal-like structure of aggregates some modifications to the equations of motion (in form of modified parameters) will be introduced in the next section. While this represents only a very simplified model and the surface forces acting on particles with a complex structure are an extremely complex problem where to date no satisfying expressions exists, we believe that this is a reasonable starting point. On the one hand, if one wants to employ the model discussed here to a specific case where better expressions are known, this can easily be adapted without changing the general idea of our approach. On the other hand, it has been found in many cases (see for example \cite{Zahnow2008_2,Zahnow2008_3}) that changes in the motion of the individual particles usually do not lead to significant changes in the dynamics of the particle ensemble and in particular in the collision rates which are relevant for the overall size distribution in an aggregation-fragmentation system.\\
Finite-size particles usually do not follow exactly the motion of the surrounding fluid, instead inertia effects lead to deviations of the particle motion from that of the fluid. For small particle Reynolds numbers the equations of motion for spherical particles of finite size are the Maxey-Riley equations \cite{Maxey1983}. This implies that locally the flow around the particle is laminar, even though the overall fluid flow can still be turbulent. While inertia effects can be fairly small for the primary particles in the case of marine aggregates (see section \ref{sec:implementation}), the influence of particle inertia increases with aggregate size and can become quite important for larger marine aggregates. \\
In reality every particle produces perturbations in the flow that decay inversely proportional to the distance from the particle \cite{Happel}. In this work we keep the particle concentration $n$ low enough to be in a diluted regime. For particles of radius $r$ and density $\rho_p$ moving coherently within the dissipative scale $l_p$ of a flow the feedback from the particle motion on the flow can be neglected if $nrl_p^2\ll1$ \cite{Balkovsky2001}. Particle-particle interactions mediated by flow perturbations are neglected, see the discussion in section \ref{sec:Limitations}. \\
Assuming that the difference between the particle velocity ${\bf V}(t)$ and the fluid velocity  ${\bf u}={\bf u}({\bf X}(t),t)$ at the position ${\bf X}(t)=(X_1(t),X_2(t),X_3(t))$ of the particle is sufficiently small, the drag force is proportional to this difference. This is called Stokes drag. With these restrictions the dimensionless form of the governing equation for the path ${\bf X}(t)$ of such a particle under the influence of drag and gravity can then be approximated from the Maxey-Riley equations as:
\beq
\dot{\bf V}=\frac{1}{St}\left({\bf u}-{\bf V}-W{\bf n}\right), 
\label{eq:maxey}
\eeq
where ${\bf n}$ is the unit vector pointing upwards in the vertical direction (which is the $X_2$-axis in this study). \\ 
The particle Stokes number $St$, i.e. the ratio between particle response time and flow time scale is defined as
\beq
  St=(\rho_P 2r^2)/(9\mu \tau_f) 
\eeq
and the dimensionless settling velocity in a medium at rest $W$ is defined as
\beq
 W=2 r^2 (\rho_P-\rho_F) \tau_f g/(9 \mu l_f)~.
\eeq
Here, $\rho_F$ and $\mu$ are the fluids' density and dynamic viscosity and $l_f$ and $\tau_f$ are characteristic length and time scales of the flow.

% --------------------------- Fractal-like aggregates ---------------------------------------------------------------
\subsection{Fractal-like aggregates}
When looking at real aggregates they are typically not solid spherical particles but instead can have a complex structure as they consist of a number of primary particles. In this model the primary particles are assumed to be solid spherical particles, following the equations of motion as described in the last subsection. All aggregates are assumed to consist of an integer number of such spherical primary particles. The description of the motion of an aggregate with a complex structure is still an unsolved problem. Therefore we only consider the influence of the structure of the aggregates on their size and effective density. Here, we assume that aggregates have a fractal-like structure, meaning that there exists a power-law relationship between the characteristic length and the mass of such aggregates. The structure of the aggregates can then be characterized by a fractal dimension $d_f<3$ \cite{Mandelbrot1983}. Their size can still be defined approximately by a radius, that can be considered as the characteristic length scale of the aggregate. This radius $r_{\alpha}$ of an aggregate that consists of $\alpha$ primary particles and has a given fractal dimension $d_f$ is derived in the following. We emphasize that the number of primary particles $\alpha$ in an aggregate is here also used as an index to describe a quantity, e.g. the radius or the volume, that corresponds to an aggregate consisting of $\alpha$ primary particles.\\
The solid volume, i.e. the volume of an aggregate that is filled with particulate matter follows from the definition of the fractal dimension $d_f$ (see e.g. \cite{Logan}) as
\beq
\label{eq:solid1}
V_{\alpha,\text{solid}}=c_{d_f}r_{\alpha}^{d_f}~,
\eeq
where $c_{d_f}$ is a proportionality constant that can depend on $d_f$. As mass conservation must be fulfilled we get
\beq\label{eq:solid2}
V_{\alpha,\text{solid}}=\alpha V_{1} = \alpha \frac{4}{3}\pi r_1^3~,
\eeq
where $r_1$ and $V_1$ are the radius and volume of a primary particle, respectively. The proportionality constant $c_{d_f}$ can be derived from Eqs. \eqref{eq:solid1} and \eqref{eq:solid2} by setting $\alpha=1$ \cite{fractality}
\beq
c_{d_f} = \frac{4}{3}\pi r_1^{3-d_f}~.
\eeq
In combination with Eqs. \eqref{eq:solid1} and \eqref{eq:solid2} this leads to 
\beq \label{eq:radius}
r_{\alpha}=\alpha^{1/d_f}r_{1}
\eeq
for the radius of an aggregate. It is evident that due to the fractal-like structure the radius $r_{\alpha}$ is greater than for a completely solid particle of the same mass. \\ 
A part of a fractal-like aggregate, i.e. of the total volume encased by an aggregate
\beq \label{eq:volume_total}
V_{\alpha,\text{total}}=\frac{4}{3}\pi r_{\alpha}^3~,
\eeq
 is not filled with matter but with the surrounding fluid. The aggregate density therefore decreases with an increasing number $\alpha$ of primary particles in the aggregate. From mass conservation it follows 
\beq\label{eq:volume}
\rho_{\alpha}V_{\alpha,\text{total}} =\rho_1V_{\alpha,\text{solid}}+\rho_F(V_{\alpha,\text{total}}-V_{\alpha,\text{solid}})
\eeq
Solving Eq. \eqref{eq:volume} for $\rho_{\alpha}$ and substituting $V_{\alpha,\text{total}}$ and $V_{\alpha,\text{solid}}$ leads to
\beq
\rho_{\alpha}=\rho_F+(\rho_1-\rho_F)\alpha^{1-3/d_f}~.
\eeq
Going back to the equations of motion we now see that as far as the particle dynamics are concerned, a first approximation for the fractal-like aggregates is to treat them as spheres with an increased radius $r_\alpha$ compared to solid spheres of the same mass, but a reduced density $\rho_\alpha$, because of the fluid encased in their spherical volume.\\ 
This leads to a modification of the particle Stokes number $St$ and dimensionless settling velocity $W$ for fractal-like aggregates in the equations of motion \eqref{eq:maxey}, when compared to a solid sphere:
\begin{eqnarray} 
St_{\alpha} &= &(\rho_\alpha 2r_\alpha^2)/(9\mu \tau_f) \\
W_{\alpha} &= & 2 r_\alpha^2 (\rho_\alpha-\rho_F) \tau_f g/(9 \mu l_f)~. 
\end{eqnarray}
For fractal-like aggregates these parameters replace $St$ and $W$ in Eq. \eqref{eq:maxey}, leading to a motion with different parameters for aggregates with different numbers $\alpha$ of primary particles.
% --------------------------- Aggregation model---------------------------------------------------------------
\subsection{Aggregation model}
The physical, chemical or biological process that leads to aggregation of particles can vary from case to case and is not examined in detail here, as this is beyond the scope of this study. Instead, a general model is used. The only assumption is that during a collision particles can somehow stick together and form an aggregate. No detailed mechanism of sticking is considered.\\
Whenever the distance between two particle centers becomes smaller than the sum of their radii, these two particles aggregate immediately, creating a new particle that replaces the two old particles. For two particles of radius $a_{\alpha_i}$ and $a_{\alpha_j}$ the new number of primary particles after aggregation is obviously $\alpha_{\text{new}}=\alpha_i+\alpha_j$, leading to a new radius $a_{\alpha_{\text{new}}}$, which can be derived from Eq. \eqref{eq:radius}. The position of the new particle is the center of gravity of the old particles. The velocity of the new particle follows from momentum conservation.

% --------------------------- Fragmentation model---------------------------------------------------------------
\subsection{Fragmentation model}
The physical process leading to fragmentation can vary as well. While there are detailed models for the fragmentation of e.g. water droplets \cite{Villermaux2007}, the mechanism of fragmentation of marine aggregates is not well understood. Even experiments give no clear indication how fragmentation of such aggregates occurs in detail. Therefore only a few theoretical approaches exist for this process \cite{Pandya1982,Hill1996}. In this work a model for fragmentation is used that is only based on some very general properties of the aggregates involved. \\
In the following the fragmentation is described in two parts, that need to be clearly distinguished. Firstly, a \textit{splitting condition}, that determines \textit{if} a fragmentation event takes place and secondly, a \textit{splitting rule}, that determines \textit{how} fragmentation takes place, are defined. \\

\subsubsection{Splitting condition}
\label{sec:split_cond}
The splitting condition describes if fragmentation of an aggregate takes place. Generally, only particles that are composed of more than one primary particle can fragment. The break-up of an aggregate occurs when the hydrodynamical forces $F_{\text{hyd}}$ acting on the aggregate exceed the forces $F_{\text{agg}}$ holding the particles in the aggregate together. The criterion for breakup can therefore be expressed as
\beq \label{eq:splitting_cond}
F_{\text{hyd}}/F_{\text{agg}}>const.
\eeq
For aggregates with a fractal-like structure, consisting of a number of solid spheres, the hydrodynamical forces in the dissipative range where viscous forces dominate is proportional to the shear force integrated over the surface of the aggregate \cite{Kobayashi1999}. For a fracture at a distance $\zeta\cdot r_{\alpha},~\zeta\in[0,1[$ from the equator of the aggregate this results in
\beq
F_{\text{hyd}} \propto S (1-\zeta) r_{\alpha}^{2}~,
\eeq
where $S$ is the shear rate in the flow.\\
For a porous aggregate Ruiz et al. \cite{Ruiz1997} give the force $F_{\text{agg}}$ holding an aggregate together as proportional to the area of constituent matter in a cross-section of the aggregate. Ruiz et al. then related $V_{\alpha,\text{solid}}$ to the porosity of an aggregate. Here, we instead rewrite this relationship in terms of the fractal dimension $d_f$. For fractures across the equator of an aggregate the area is proportional to $V_{\alpha,\text{solid}}^{2/3}\propto r_1^2\left(\frac{r_{\alpha}}{r_1}\right)^{2d_f/3}$. For a fracture at a distance $\zeta\cdot r_{\alpha},~\zeta\in[0,1]$ the area of constituent matter in the cross-section is reduced. It equals the area of a cross-section across the equator of an aggregate with decreased radius $(1-\zeta^2)^{1/2}r_{\alpha}$. In general we get
\beq
F_{\text{agg}} \propto (1-\zeta^2)^{d_f/3}r_{\alpha}^2 \left(r_{\alpha}/r_1\right)^{2d_f/3-2}~.
\eeq
The splitting condition \eqref{eq:splitting_cond} then becomes
\beq
S\cdot\frac{(1-\zeta)}{(1-\zeta^2)^{df/3}}\cdot\left(\frac{r_{\alpha}}{r_1}\right)^{2-2d_f/3}   > \gamma,
\eeq
where the proportionality constant $\gamma$ is determined by the force holding the primary particles in an aggregate together. It is therefore a measure of the aggregate strength.  Solving for the shear rate $S$ and using Eq. \eqref{eq:radius} leads to
\begin{eqnarray} \label{eq:splitting_agg}
\centering
S &> &\gamma\frac{(1-\zeta^2)^{df/3}}{(1-\zeta)} \left(r_{\alpha}/r_1\right)^{-2+2d_f/3}\nonumber \\
&= &\gamma \frac{(1-\zeta^2)^{df/3}}{(1-\zeta)} \alpha^{2/3-2/d_f}~.
\end{eqnarray}
It can be seen that for a fractal dimension $d_f<3$ the critical shear force required to break up an aggregate decreases with the aggregate size, i.e. larger aggregates are less stable than smaller ones. Additionally, the critical shear required for fragmentation is smallest for fractures across the equator of an aggregate ($\zeta=0$) and increases with increasing distance from the equator.\\
The shear force $S$ is given by 
\beq
S=\left(2\sum\limits_{i,j}S_{ij}S_{ij}\right)^{(1/2)}~,
\eeq
where $S_{ij}=\frac{1}{2}\left(\frac{\partial u_i}{\partial X_j}+\frac{\partial u_j}{\partial X_i}\right)$ is the rate-of-strain tensor in the flow.

\subsubsection{Splitting rules}
\label{sec:splitting_rules}
\begin{figure}[htb]
		\centering
		\includegraphics[width=0.3\textwidth]{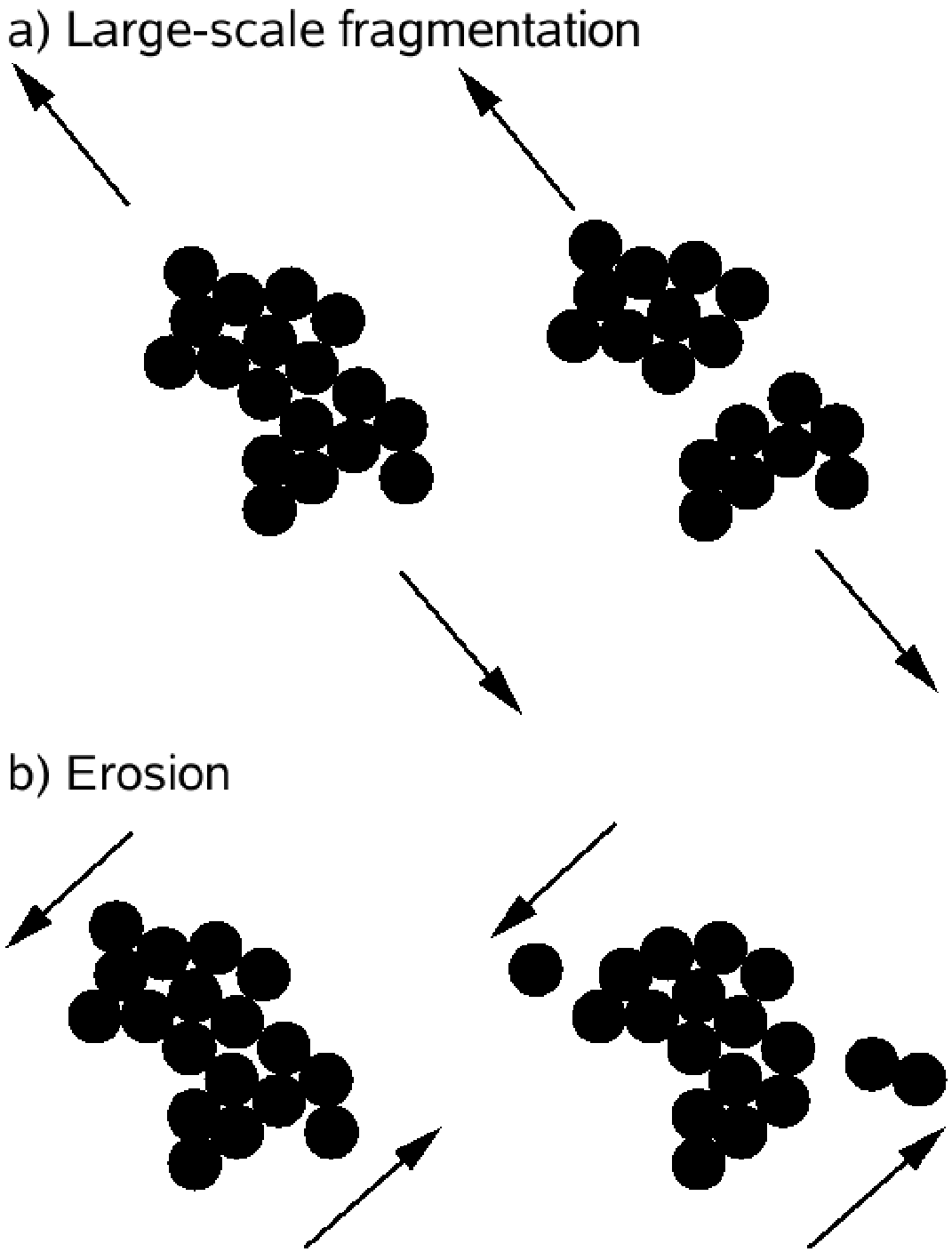}
		\caption{\label{fig:splitting_rules} Sketch of different splitting mechanisms: a) large-scale fragmentation where an aggregate is pulled apart along the equator, b) erosion where small parts are split off from the aggregate surface.}	
\end{figure}

The splitting rules describe how an aggregate will split, when the splitting condition is met. During fragmentation only particles whose mass is an integer multiple of the mass of a primary particle are created. This means that, even though they have become part of some larger aggregates, primary particles can never be broken up. Only the connection among each other can break. Different size distributions of the fragments are possible. \\
When a splitting condition is met, an aggregate consisting of $\alpha_{\text{old}}$ primary particles is split into $2$ fragments with the number $\alpha_k$ of primary particles of each fragment being a random fraction of the original number $\alpha_{\text{old}}$. \\
Typically, one distinguishes between two different mechanisms of fragmentation \cite{Jarvis2005}. For each mechanism fragmentation occurs on average at a different distance from the equator of the aggregate, leading to different distributions of the fragments $\alpha_k$. This can be expressed in different values for $\zeta$, the fraction of the distance from the equator of the aggregate where fragmentation is assumed to take place (see Sec. \ref{sec:split_cond}). \textit{Large-scale fragmentation} happens when an aggregate is 'pulled apart' somewhere close to the equator, leading to fragments of similar size. This is characterized by $\zeta=0$. \textit{Erosion} happens when shear forces act closer to the edge of an aggregate \cite{Vassileva2007}, implying that $0< \zeta <1$. In this case only few primary particles are split off from the aggregate (see Fig. \ref{fig:splitting_rules}).\\
Here we will compare three different fragmentation modes, with different distributions for the number of primary particles in the fragments. First a large-scale splitting rule, second an erosion splitting rule, and third a \textit{uniform splitting} rule. In most realistic cases one expects different fragmentation modes to appear together, even though with slightly different probabilities. However, here we apply these fragmentation modes separately to determine their individual influence on the aggregate size distribution.

\begin{enumerate}
 \item For both large-scale fragmentation and erosion we assume that there is a preferred distance $\zeta r_\alpha,~\zeta\in[0,1[$ from the center of the original aggregate where fragmentation occurs. The two mechanisms are then characterized by different values of $\zeta$. A fracture at a distance $\zeta r_\alpha$ leads to a fragment containing a fraction $V_{\text{fr}}/V_{\alpha,\text{total}}$ of the original volume, where $V_{\text{fr}}=\pi r_\alpha^3(\frac{2}{3}-\zeta+\frac{\zeta^3}{3})$. Since $\zeta$ is assumed to be only the average distance of a fracture from the equator of the aggregate, for each fragmentation event we choose the number of primary particles in the fragment randomly from a Gaussian distribution, centered around $\alpha_{\text{old}}\cdot V_{\text{fr}}/V_{\alpha,\text{total}}$. This allows for a certain variation of the fragment size, meaning that
\beq \label{eq:same_splitting}
\alpha_{1}=\left(\frac{4}{3} \left(\frac{2}{3}-\zeta+\frac{\zeta^3}{3}\right)+\xi\right)\alpha_{\text{old}}~,
\eeq
where $\xi$ is a random number from a normal distribution with mean $0$ and standard deviation $\sigma_F$ and the right-hand side of equation \eqref{eq:same_splitting} is rounded towards the nearest integer. We note that our results do not depend strongly on the choice of $\sigma_F$, here we choose $\sigma_F=0.2$ which results in a typical variation of one primary particle between fragments coming from identical aggregates. As an additional restriction it is required that $1\leq\alpha_{1}<\alpha_{\text{old}}$, otherwise a new random number $\xi$ is chosen.\\
 For large-scale fragmentation, aggregates are assumed to break along the equator into two fragments of similar size, which corresponds to $\zeta=0$. For erosion, fragmentation is assumed to occur at a distance from the center, leading to one aggregate being much smaller than the other. This corresponds to $0<\zeta<1$. Here, we choose $\zeta=0.6$ for erosion, which leads to smaller fragments containing on average 10 percent of the mass of the original aggregate. Similar results for erosion were found for other values of $\zeta$. However, if $\zeta$ becomes too large particles will no longer fragment because the critical shear required to break off a fragment increases greatly as $\zeta\rightarrow1$. 
\item In the uniform splitting rule the number of primary particles for the first fragment is chosen from a uniform distribution in the interval $I=[1,\alpha_{\text{old}}[$. The uniform splitting rule is a simplified model for the full case where both large-scale fragmentation and erosion of an aggregate can happen. However, all sizes of fragments occur here with the same probability. 
\end{enumerate}
For all three cases the remaining aggregate consists of $\alpha_2=\alpha_{\text{old}}-\alpha_1$ primary particles.\\
Whenever a particle is split according to one of the splitting rules, all parts keep the velocity of the original particle. That way momentum is conserved. The first fragment remains at the position ${\bf X}=(X_{1},X_{2},X_{3})$ of the original particle. The center of the other fragment is placed along a line segment in a random direction, so that for the two fragments the distance equals the sum of the radii.\\
For each fragment, the splitting condition is checked again and if it is met, the whole process is repeated until no fragment fulfills the fragmentation condition. This leads to a splitting cascade and aggregates can break up into more than two fragments, if the aggregate is large enough or shear forces are strong enough. This is consistent with experimental observations of marine aggregates that larger particles tend to break into more fragments \cite{Alldredge1990}. In that way ternary, quarternary and other splitting types besides binary splitting naturally appear in this model.\\
Here, large-scale fragmentation and erosion are treated as two separate processes to study the influence of fragmentation at certain distances $\zeta$ on the aggregate size distributions in the steady state. In reality, aggregates will break with certain probabilities at certain distances $\zeta$ from the center but there will be no two separate processes. Therefore, depending on the probability distribution for fragmentation at a certain distance one can expect different combinations of the steady state size distributions found in this work. The uniform fragmentation rule represents one such possible combination, where fragmentation at all distances $\zeta$ appears with the same probability.

% --------------------------- Fluid Flow ---------------------------------------------------------------
\subsection{Fluid Flow}
\label{sec:Fluid Flow}
As a fluid velocity field we consider synthetic turbulence in the form of a space-periodic, isotropic and homogeneous Gaussian random flow \cite{Bec2005_2}, since it allows us to perform long-term simulations at reasonable computational costs.  We use a smooth, incompressible flow since we focus on effects typically taking place on scales smaller than the Kolmogorov scale of a turbulent flow. 

The flow is written as a Fourier series
\begin{equation}\label{eq:A1_Fouriersum}
\vec{u}(\vec{X},t) = \sum_{\vec{k}\in\mathbb{Z}^d\backslash\{\vec{0}\}} \vec{\hat{u}}(\vec{k},t)e^{i\frac{2\pi}{L}\vec{k}\cdot \vec{x}}~,
\end{equation}
where $\vec{\hat{u}}(\vec{k},t)\in\mathbb{C}^d$ are the Fourier components, with the property $\vec{\hat{u}}(-\vec{k},t)=\vec{\hat{u}}^*(\vec{k},t)$ because $\vec{u}(\vec{X},t)$ is real-valued. The star denotes complex conjugation. By taking for $\vec{\hat{u}}(\vec{k},t)$ the projection of a different vector $\vec{\hat{v}}(\vec{k},t)\in\mathbb{C}^d$ onto the plane perpendicular to the wave vector $\vec{k}$, incompressibility is ensured. The vector $\vec{\hat{v}}(\vec{k},t)$ is assumed to be an Ornstein-Uhlenbeck process. It is a solution of the complex-valued stochastic differential equation 
\begin{equation}
d\vec{\hat{v}}= -\xi(\vec{k})\vec{\hat{v}}dt+\sigma(\vec{k})\vec{dW}~,
\end{equation}
with $\xi(\vec{k}),\sigma(\vec{k})\in\mathbb{R}$, where $\vec{dW}$ is a $d$ dimensional complex Wiener increment. The parameters $\xi(\vec{k}),\sigma(\vec{k})$ need to be chosen in such a way that the flow $\vec{u}(\vec{x},t)$ reproduces some features of a real turbulent flow, in this case we choose the energy spectrum in the dissipative range of a turbulent flow. Here we use the exponential spectrum suggested by Kraichnan 
\begin{equation}\label{eq:spec_kraichnan}
E(k) = C\cdot(2\pi kl_f/L)^3\exp(-\beta [2\pi kl_f/L])~,
\end{equation} 
with $\beta=5.2$ \cite{Martinez1997} and a suitably chosen normalization constant $C$. The constant $l_f$ is the length scale of coherent structures in the flow and $L$ is the spatial period of the flow. We choose $\xi(k)=1/\tau_f$ and $\sigma(k)=\sqrt{E(k)/\tau_f}$. The constant $\tau_f$ is then the correlation time of the flow. The normalization constant is chosen so that a desired value of the mean shear rate $\left<S\right>$ is obtained. The flow is then characterized by the correlation time $\tau_f$, the correlation length $l_f$ and the mean shear rate $\left<S\right>$. 

If a fluid velocity field with few Fourier modes is chosen, no interpolation of the velocity at particle position is required, since it can be calculated from direct summation of the Fourier series. This allows for a resolution of the fine structures of the particle distribution in space.

\subsection{Implementation}
\label{sec:implementation}
Next, we will describe some specifics of the numerical implementation and the system parameters used in this work.\\
For the simulations in this work we choose particle properties similar to those of marine aggregates in coastal waters. The primary particles considered in this model have a radius $r_{1}=4\mu$m, density $\rho_1=2.5\times10^3\text{kg}/\text{m}^3$ and mass $m_1=\rho_1 \frac{4}{3}\pi r_1^3$. The relevant characteristic length scales for marine aggregates in coastal areas of the ocean are typically the Kolmogorov scales. Shear rates in coastal areas can be of order $1s^{-1}$, leading to Kolmogorov length and time scales of $l_f=1$mm and $\tau_f=1$s, respectively \cite{Kranenburg1994}. Using these scales to make the equations of motion of the particles dimensionless leads to a Stokes parameter of $St_1=10^{-5}$ and a dimensionless settling velocity of $W_1=0.1$ for the primary particles. \\
The aggregate strength parameter $\gamma$ is fixed at $\gamma=8$, unless otherwise mentioned.\\
The number of aggregates $N(t)$ changes over time due to aggregation and fragmentation leading to a distribution of aggregates of different radii in the flow. However, the total mass $M=\sum_{i=1}^{N(t)}\alpha_{i}m_{1}$ remains constant during one simulation. As initial condition we take $10^5$ primary particles and no larger particles. Furthermore particles are uniformly distributed over one periodic cell of size $L^3$ of the configuration space, with velocities matching that of the fluid. This choice fixes the total mass of the system to be $M=10^5 m_1$. For the flow we choose a periodic cube with $L=4l_f$, so that we obtain a volume fraction of about $0.4\times10^{-3}$.\\
The fractal dimension $d_f$ of marine aggregates varies between approximately $1.5$ for very open, fragile aggregates like marine snow in the open ocean and approximately $2.5$ for stronger, compact aggregates. The average is typically around $1.9-2.0$ (see e.g. \cite{Winterwerp1998}), therefore in the following we choose $d_f=2.0$ unless otherwise mentioned.\\
As a first approximation the aggregation and fragmentation processes are assumed to have no effect other than to change the size of the particles and the effective density, and therefore do not directly influence the motion of the particles. Hence all three aspects, motion, aggregation and fragmentation that define the whole system can be modeled separately. Aggregation is checked constantly during the integration, whereas fragmentation is applied after every time step of the system. 

\begin{enumerate}
 \item All particles move in the flow for one time step $dt$ using the equations of motion described in Sec. \ref{sec:Equations}. We emphasize at this point that each aggregate size is characterized by different values of $St_{\alpha}$ and $W_{\alpha}$, so that the motion of aggregates of different size is governed by the same equations but with different parameters. \\
The length of the time step $dt$ needs to be chosen small enough so that the simulation result becomes independent of this values, here we found $dt=0.01$s to be sufficiently small. \\
Because of the spatial periodicity of the flow, all particle dynamics will be folded back onto one $L^3$ cell in the flow, using periodic boundary conditions. Usually particles do not stay in one cell of the flow, i.e. they are not suspended in the flow. Instead particles will generally fall downwards through the flow, if they are heavier than the fluid \cite{Maxey1986}. This means that folding the dynamics of the particles back onto a single cell is only a convenient way to visualize an infinitely extended system and does not completely mirror what one would see in a comparable experiment. However, if particles are initially distributed homogeneously over the whole configuration space, the total particle mass in each periodic cell remains the same over time. Therefore even if aggregation and fragmentation are included, it is sufficient to restrict our studies to one unit cell with periodic boundaries.
\item Particles aggregate upon collision, i.e. if their distance becomes equal to the sum of their radii. To ensure that no collisions are missed, we use an efficient event-driven algorithm for particle laden flows (cf. \citet{Sigurgeirsson2001} for details). Computationally, the aggregation process is the most costly component of the simulation. In particular, the naive approach to check which particles are colliding involves looping over all pairs of particles and therefore scales as $O(N^2)$, where $N$ is the number of particles. Therefore, here a linked-list algorithm \cite{Hockney1981}, sometimes also called link-cell algorithm is used to compute the distance between particles. The configuration space is divided into grid cells of size $\epsilon$, where each grid cell stores information on which particles it contains. The looping over particle pairs to calculate their distance is then done only over particles in a given grid cell and the neighboring cells. If the grid cell size $\epsilon$ is small enough (but larger than the largest appearing particle size) the link-cell algorithm scales as $O(N)$ and is thus much faster than the naive approach. 
\item After each time step $dt$ particles can fragment if the shear at their position exceeds a critical value. If that is the case, new fragments are created according to the rules described in section \ref{sec:splitting_rules}.
\end{enumerate}

We note that at first glance it looks like aggregation and fragmentation are treated very different, in particular aggregation seems to be independent of the aggregate strength $\gamma$ in this model. However, this is not the case. Initially all particles that come into contact aggregate, i.e. here the aggregation probability upon collision is equal to one. But when looking at aggregation and fragmentation together over one time step $dt$ it is in fact smaller than one because some aggregates that just formed during this time step will break up again. These are the aggregates where the aggregate strength $\gamma$ is not strong enough to hold the aggregate together. This means after one time step only some of the particles that came into contact will actually have aggregated and this number will depend on the aggregate strength $\gamma$. This means that both aggregation and fragmentation probabilities depend on the same aggregate property, which one would expect in reality.\\

% ---------------------------------------------------------------------------------------------------------------
%					RESULTS
% --------------------------------------------------------------------------------------------------------------
\section{Simulation results}
\label{sec:Results}
In the following section we will present simulation results using the model described above, to determine the influence of the different splitting rules on the resulting particle size distributions. As the parameters used in the model system can vary greatly in natural systems, we examine the sensitivity of the system with regard to the following parameters: aggregate strength $\gamma$, fractal dimension $d_f$ and total number of primary particle $N$.\\

\subsection{Measured quantities}
From previous works it is known that the balance of aggregation and fragmentation typically leads to a steady state of the particle size distribution \cite{Zahnow2009_2}. This follows from the fact that normally aggregation dominates for small sizes, whereas fragmentation is the dominant process for large sizes. In addition to studying this size distribution of the particles in the steady state, we introduce different measures to characterize first the approach to the steady state and then the steady state itself. \\ 
To follow the convergence of the system towards a steady state, we use two different quantities. The first quantity that we measure during the simulations is the average number of primary particles per aggregate, defined as
\beq
\left<\alpha(t)\right> = \sum_{\alpha}\alpha N_{\alpha}(t)/N(t)~.
\eeq 
In the context of our model $2\left<\alpha(t)\right>$ corresponds to the 'mean equivalent circular diameter' that is often used as a measure in experiments with marine aggregates \cite{Lunau2006}. We will use this quantity as a first estimate whether the particle size distribution has converged to a steady state and to follow the evolution of the particle size distribution towards the steady state. \\
A second quantity of the aggregation and fragmentation process that may be experimentally measured is the time it takes to reach the steady state. Especially in technical applications this can be an important quantity, where processes need to be timed appropriately to allow for a smooth work flow. 
Here we introduce a measure for this relaxation time in our model and show how different system parameters influence this time to reach the steady state. \\
We define the relaxation time $\tau_\infty$ as 
\beq\label{eq:relaxtime}
 \tau_\infty=\int_{0}^{\infty}dt\left(\left|1-\overline{\left<\alpha(t)\right>}/\alpha_\infty\right|\right)~.
\eeq
$\overline{\left<\alpha(t)\right>}$ is a running (time)-average of the average number of primary particles in an aggregate. It is used to remove fluctuations due to the periodic changes in the flow. This definition of the relaxation time is analogous to the definition of the correlation time for stochastic processes as the integral over the autocorrelation function. For example, for an exponential relaxation process $\propto e^{-t/t_R}$ Eq. \eqref{eq:relaxtime} leads to the expected result of $\tau_\infty=t_R$. However, Eq. \eqref{eq:relaxtime} stays an appropriate measure for more irregular relaxation processes.\\
As a simple measure to characterize the steady state of the particle size distribution we use the average number of primary particles in an aggregate in the steady state that is defined as
\beq
\alpha_\infty=\lim\limits_{t\rightarrow\infty}\left<\alpha(t)\right>~.
\eeq 

% --------------------------- Approach to a steady state ---------------------------------------------------------------
\subsection{Approach to a steady state}
\begin{figure}[htb]
		\centering
		\includegraphics{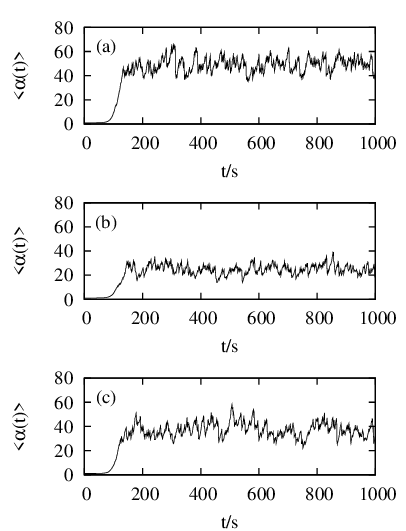}
		\caption{\label{fig:steady_state} Average number of primary particles as a function of time for aggregate strength $\gamma=9$ and aggregate fractal dimension $d_f=2.0$ for (a) large-scale fragmentation (b) erosion and (c) uniform fragmentation. } 	
\end{figure}
First, we use the average number of primary particles per aggregate to follow the convergence of the systems towards a steady state for all three splitting rules (Fig. \ref{fig:steady_state}). Our initial condition is always a uniform distribution of primary particles.\\ 
Initially, aggregation leads to a fast increase in the average number of primary particles per aggregate similar for all splitting rules. 
\begin{figure}[htb]
		\centering
		\includegraphics{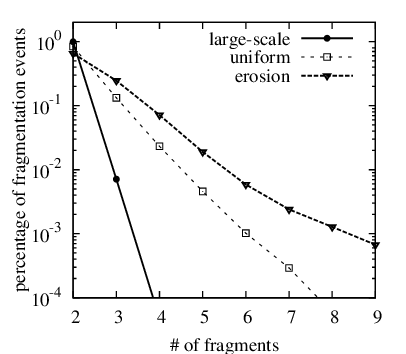}
		\caption{\label{fig:breaking_probability} The histogram shows the percentage of fragmentation events for the number of fragments created in that event, for the same simulation run as shown in Fig. \ref{fig:steady_state}. Large-scale fragmentation leads typically to the smallest number of fragments, while erosion typically generates most fragments.} 	
\end{figure}
 Then fragmentation sets in and a balance between aggregation and fragmentation is reached, with a different steady state average particle size  for the different splitting rules. $\alpha_{\infty}$ fluctuates over time, due to the statistical fluctuations in the flow.  Large-scale splitting leads to the highest average number of primary particles per aggregate, erosion to the lowest and uniform fragmentation is in between. This can be intuitively understood, since for erosion typically more fragments are created than for large-scale fragmentation (see Fig. \ref{fig:breaking_probability}). When a particle gets eroded, one of the fragments is usually close to the same size as the original aggregate. This leads to a high probability that this fragment will break again and therefore in many cases fragmentation will not be binary, but many fragments will be created. For large-scale fragmentation, aggregates will typically break only once, since both fragments are much smaller than the original aggregate. \\
In general, it is less likely in the case of large-scale fragmentation that a large number of fragments is created. This leads in the mean to a larger average aggregate size than for erosion.\\
The different splitting rules lead to very different distributions (cf. Fig. \ref{fig:aggstrength_dist}). Large-scale fragmentation creates a distribution with a single peak at intermediate radii and no particles of the smallest sizes. The right hand side of the aggregate size distribution for large-scale fragmentation follows approximately an exponential decay. The size distribution found for large-scale fragmentation corresponds well to those observed for marine aggregates \cite{Lunau2006} where exponential size distributions have also been reported.  This may indicate that large-scale fragmentation is indeed the primary mode of break-up for many marine aggregates, as proposed in some works (see e.g. Ref. \cite{Thomas1999}) and that erosion plays a very small role there. \\
By contrast, the size distribution for erosion has two different regimes, with a sharp maximum at the smallest aggregate size and a slower decaying tail at larger aggregate sizes.\\
Uniform splitting, where both larger and smaller fragments are created leads to a plateau in the size distribution at smaller aggregate sizes and an exponential decay towards larger aggregate sizes. \\
Many of the system parameters that appear in our model can vary much in natural systems, in particular the aggregate strength, the number of primary particles involved and the fractal dimension of aggregates. In the following we therefore examine the sensitivity of our results to these parameters.

% --------------------------- Aggregate strength---------------------------------------------------------------
\subsection{Influence of aggregate strength}
\begin{figure}[H]
		\centering
		\includegraphics{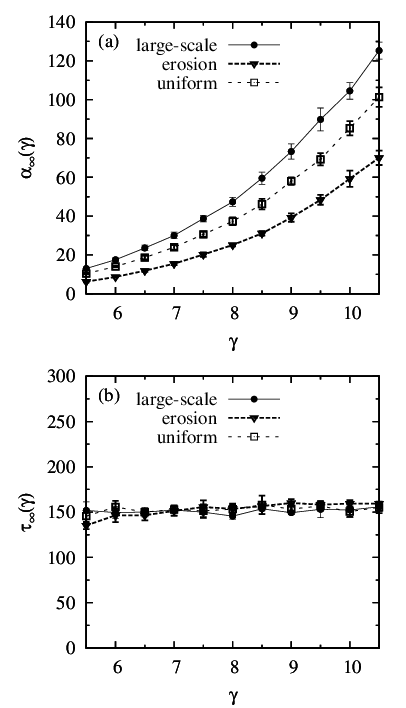}
		\caption{\label{fig:aggstrength} Influence of aggregate strength $\gamma$. (a) Average number of primary particles per aggregate in steady state $\alpha_\infty$  and (b) relaxation time $\tau_\infty$ for the approach to the steady state for different values of the aggregate strength $\gamma$. Error bars are obtained from an ensemble of $5$ different realizations of the carrier flow.} 	
\end{figure}

To determine the influence of the forces holding the aggregates together on the resulting steady state size distribution, $\alpha_\infty$ is computed for different values of the aggregate strength $\gamma$. $\alpha_\infty$ increases with $\gamma$ for all fragmentation rules. The increase is fastest for large-scale fragmentation and slowest for erosion. (see Fig. \ref{fig:aggstrength}(a)). 
The relaxation time $\tau_\infty$ as a function of $\gamma$ is shown in Fig. \ref{fig:aggstrength}(b). For all three fragmentation rules, the relaxation time is independent of the value of $\gamma$. 
\begin{figure}[H]
		\centering
		\includegraphics{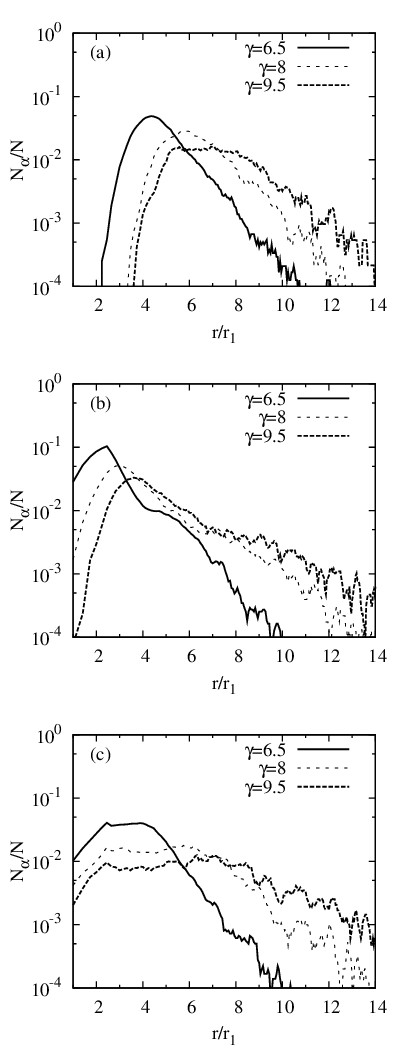}
		\caption{\label{fig:aggstrength_dist} Histogram of the particle size distribution. Number of particles versus the  radius $a$ for different values of $\gamma$ for (a) large-scale splitting (b) erosion splitting and (c) uniform splitting.}
\end{figure}
The relaxation time is defined \textit{relative} to the average number of primary particles in steady state $\alpha_\infty$ and therefore the actual value of $\alpha_\infty$ does not influence the relaxation time.  \\
When looking at the particle size distribution in steady state (Fig. \ref{fig:aggstrength_dist}), the difference between the three fragmentation rules is again clearly visible. However, this difference does not seem to depend on the value of the aggregate strength $\gamma$, as the distributions for each fragmentation rule remain qualitatively the same for different $\gamma$, but getting wider with increasing aggregate strength. 
% --------------------------- Volume fraction---------------------------------------------------------------
\subsection{Influence of the volume fraction}
\begin{figure}[H]
		\centering
		\includegraphics{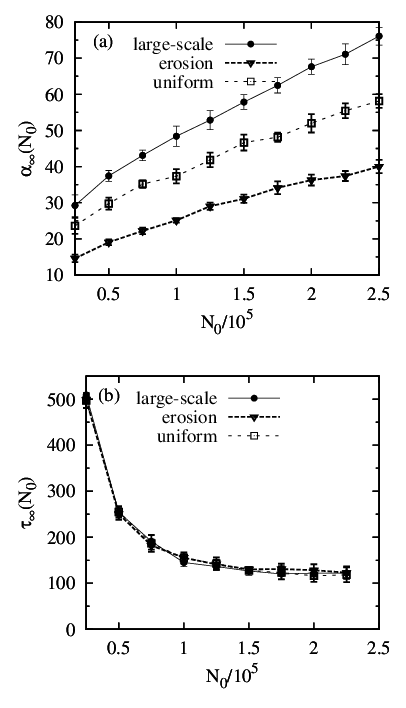}
		\caption{\label{fig:volumefrac} Influence of the total number of primary particles $N_0$. (a) Average number of primary particles per aggregate in steady state $\alpha_\infty$  and (b) relaxation time $\tau_\infty$ for the approach to the steady state for different values of the totalnumber of primary particles $N_0$. Error bars are obtained from an ensemble of $5$ different realizations of the carrier flow.} 	
\end{figure}
\begin{figure}[H]
		\centering
		\includegraphics{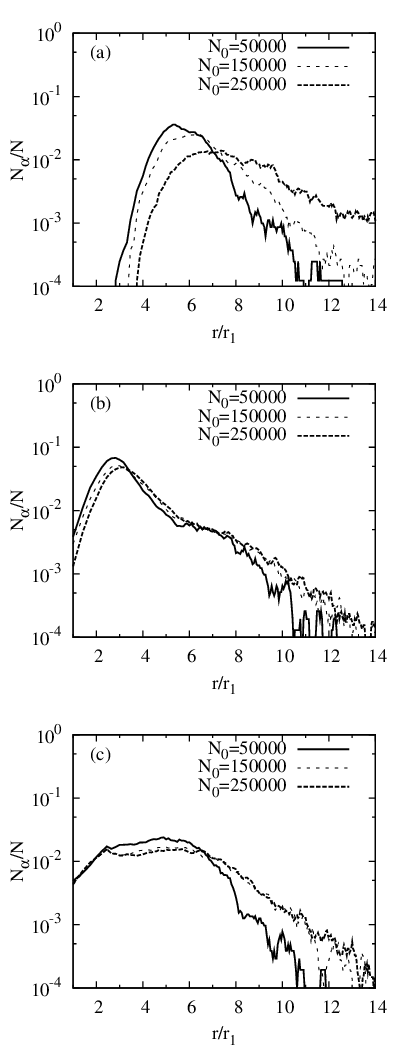}
		\caption{\label{fig:volumefrac_dist} Histogram of the particle size distribution. Number of particles versus the radius $a$ for different values of the total number of primary particles $N_0$ for (a) large-scale splitting (b) erosion splitting and (c) uniform splitting. } 	
\end{figure}
To determine the influence of the volume fraction, i.e. the total number of primary particles $N_0$ in the system, on the resulting size distribution, $\alpha_\infty$ is computed for different values of $N_0$. $\alpha_\infty$ increases with increasing number of primary particles, due to the increased number of collisions. Again, the average number of primary particles per aggregate in the steady state $\alpha_\infty$ is largest for large-scale fragmentation and smallest for erosion. The increase in $\alpha_\infty$ with $N_0$ is almost linear for all three fragmentation rules.  \\
The relaxation time (see Fig. \ref{fig:volumefrac}(b)) decreases for increasing $M$ for all three fragmentation rules. This again shows that the relaxation time does not depend strongly on the absolute value of $\alpha_{\infty}$. Instead, this indicates that the relaxation time is mainly determined by the collision rate between the particles. While a change in fragmentation rate, for example due to increased $\gamma$ dues not affect the relaxation time, an increased collision rate, due to an increased number of primary particles in the system seems to decrease the relaxation time significantly.\\
Again, the size distributions (Fig. \ref{fig:volumefrac_dist}) remain clearly different for the different fragmentation rules, independent of the total number of primary particles $N_0$ in the system.

% --------------------------- Fractal-like dimension---------------------------------------------------------------
\subsection{Influence of the fractal dimension}\label{sec:fractality}
\begin{figure}[H]
		\centering
		\includegraphics{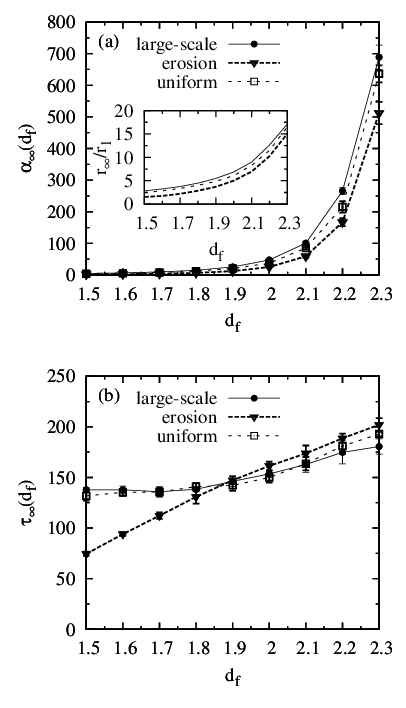}
		\caption{\label{fig:fractdim}  Influence of the fractal dimension $d_f$ of the aggregates. (a) Average number of primary particles per aggregate in steady state $\alpha_\infty$, the inset shows the relative average size of aggregates in the steady state $r_\infty/r_1=\alpha_\infty^{1/d_f}$ of the aggregates as a function of $d_f$  and (b) relaxation time $\tau_\infty$ for the approach to the steady state for different values of the fractal dimension $d_f$ of the aggregates. Error bars are obtained from an ensemble of $5$ different realizations of the carrier flow.} 	
\end{figure}

The last important system parameter that typically varies a lot in natural systems is the fractal dimension $d_f$ of the aggregates \cite{Winterwerp1998}. To determine the influence of the fractal dimension of the aggregates, $\alpha_\infty$ is computed for different values of $d_f$. For all three fragmentation rules we find a drastic increase in the average number of primary particles per aggregate in the steady state (see Fig. \ref{fig:fractdim}(a)) with increasing $d_f$. In the case of varying the fractal dimension this increase in $\alpha_\infty$ is much more drastic than for the other parameters studied in the previous sections.  $\alpha_\infty$ increases by approximately a factor of $100$ between $d_f=1.5$ and $d_f=2.3$. Initially, one might assume that the increase of the average number of primary particles per aggregate in the steady state is only due to the aggregates becoming more compact as the fractal dimension is increased and does not really reflect a change in the aggregate size. However, plotting the average radius of the aggregates in the steady state as a function of $d_f$ (inset in Fig. \ref{fig:fractdim}(a)) shows that there is also a significant increase in the aggregate size with increasing $d_f$.\\
This increase when varying the fractal dimension can be understood by looking at the stability condition for the aggregates (cf. Eq. \eqref{eq:splitting_agg}). The stability curve defined by Eq. \eqref{eq:splitting_agg} becomes almost horizontal for larger aggregate sizes. The larger $d_f$ the greater becomes the range of aggregate sizes where increasing the size has almost no effect on the stability (with the limit of $d_f=3$ where stability is independent of the size). Increasing the aggregate strength $\gamma$ also leads to larger aggregates being stable at a given value of shear force. However, the increase in the range of stable aggregate sizes is much lower.  \\
We note that the relaxation time increases weakly with $d_f$ for all splitting rules (see Fig. \ref{fig:fractdim}(b)). Increasing the fractal dimension leads to more compact aggregates, i.e. less overall volume occupied by aggregates and therefore smaller collision probabilities. This in turn increases the relaxation time. This effect appears strongest for erosion, where the relaxation time becomes very short for small values of $d_f$ whereas for large-scale fragmentation and uniform fragmentation there seems to be a saturation of the relaxation time for values of $d_f<2$. \\
Once again, the shapes of the particle size distributions retain their characteristic differences for the different fragmentation rules. Aside from the increasing fluctuations in the distributions for increasing $d_f$, due to the decreasing number of aggregates in the system, the qualitative shape of the distribution remains characteristic for the fragmentation rule, independent of the value of $d_f$. \\
 The range of $d_f$ that is observed in natural systems reaches even further than $d_f=2.3$, up to approximately $2.6$. However, this is not shown here, because due to the drastic increase in the average number of primary particles per aggregate only very few aggregates would remain for such high values of $d_f$ (if the other parameters remain fixed). Hence, no meaningful statistics or size distributions could be obtained. 

\begin{figure}[H]
		\centering
		\includegraphics{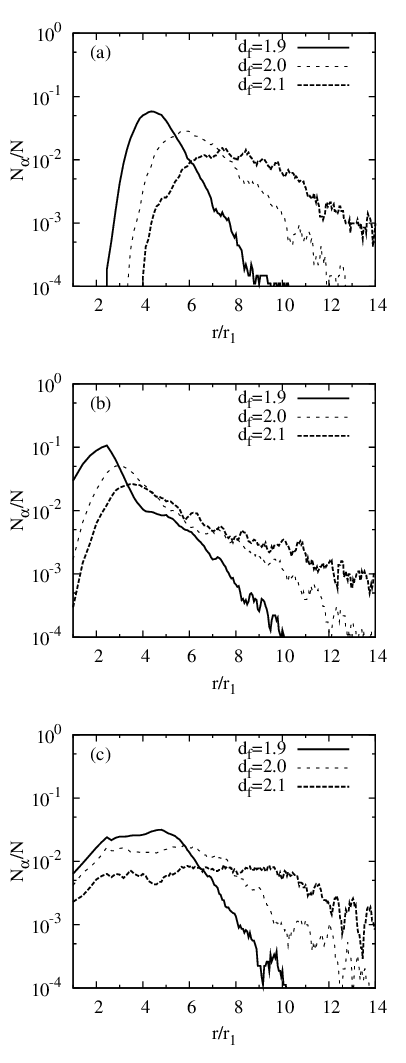}
		\caption{\label{fig:fractdim_dist} Histogram of the particle size distribution. Number of particles versus the radius $a$ for different values of the fractal dimension $d_f$ of the aggregates for (a) large-scale splitting (b) erosion splitting and (c) uniform splitting. } 	
\end{figure}
This is a computational limitation of the current model, due to the finite number of primary particles that can be studied. While it occurs for every parameter, e.g. increasing the aggregate strength $\gamma$ too far has the same effect, it is most pronounced for the fractal dimension. Small increases of the fractal dimension lead to proportionally much larger increases in the mean aggregate size, therefore reducing the total number of aggregates in the system very quickly. 

% ---------------------------------------------------------------------------------------------------------------
%				CONCLUSIONS
% ---------------------------------------------------------------------------------------------------------------
\section{Discussion and Conclusions}
\label{sec:Conclusions}
To conclude this work, in the final section we provide a brief discussion of the limitations of the particle-based aggregation and fragmentation model proposed here and finish with a brief summary of our results. 

\subsection{Limitations of the Model}
\label{sec:Limitations}
In this work we showed the application of our particle-based model to the problem of aggregation and fragmentation of marine aggregates. We emphasize that the particle-based model introduced here is a very general model that can be applied to a wide range of other problems, for example in chemical engineering and has the potential to be a useful addition to the usual modelling approaches for aggregation and fragmentation. However, there are a number of limitations that should be pointed out. Some of these are due to physical aspects of the problem that are not yet fully understood and can therefore not be captured, others are mainly due to computational limitations. \\
The main physical aspect of the problem that is still not fully understood are the details of the fragmentation mechanism. In particular for marine aggregates, but also for many other systems where aggregates with a fractal-like structure appear, there still exists no satisfying microscopic theory for the fragmentation process. The fragmentation model used in this work can therefore only be considered as a very simplified view on the problem and most likely does not capture many aspects of the real situation. However, it serves as a useful basis to consider the qualitative impact of different fragmentation mechanisms on the overall size distribution of the aggregates.\\
Additionally, the equations of motion for fractal-like aggregates can be expected to be very different from the simple equations used here. The inclusion of the increased radius and effective density which was done in this work represents a very simple modification of the relevant forces, such as the drag forces acting on the particle and is unlikely to capture the full complexity of the problem. However, it is a useful approximation to consider the impact of a complex structure on the steady state size distribution.  \\
Furthermore, the description of marine aggregates as having a fractal-like structure is only a first approximation. Measurements have shown that when averaged over many different individual aggregates there exists a power-law relationship between the aggregate size and mass. This is certainly not true for each individual aggregate, but has been found to be a reasonable model in many cases \cite{Kranenburg1994}. In addition, the concept of a fractal dimension is usually only valid over a certain size range of aggregates. Aggregates that consist of only one or two primary particles generally do not have the same structure as a larger aggregate. In the context of this work it is possible to consider as primary particles the smallest fraction of an aggregate that still has the same power-law relationship as the large aggregates, i.e. as the fractal generator of the aggregate (see also the comment \cite{fractality}).\\
In addition, there are a number of aspects that are theoretically understood quite well, but can not be included in such a model due to computational limitations. First among these is the two- and three-way coupling between the particles and the surrounding fluid. Two-way coupling, i.e the feedback of the particles on the fluid can in principle be included but requires solving the equations of motion for the fluid together with the equations for the particles and therefore leads to a drastic increase in computational effort. In addition, as the particle radius $r$ is assumed to be small, the feedback from the particle motion on the flow is usually not significant \cite{Michaelides1997} and can therefore be neglected. The inclusion of three-way particle-flow coupling, i.e. the interaction of particles through the fluid is usually a bigger challenge. In particular at low Reynolds numbers this interaction affects the particle motion even at low particle concentrations \cite{Brady1988}. While there exist a number of interaction models, that are able to compute this three-way coupling (see e.g. \cite{Knudsen2008}) they are typically limited to systems with only a few particles because of the computational costs involved. While the modification of individual particle trajectories can be significant, the main effect for the collective aggregation and fragmentation dynamics of an ensemble of particle is typically a reduction of the collision rates. This can be approximated in a simple way by introducing a collision efficiency, i.e. a probability for aggregation upon collision. However, the introduction of such a collision efficiency does not qualitatively affect the results shown here. Additionally, the fluiddynamic interaction of permeable particles such as particles with a fractal-like structure discussed here are not understood very well. It is likely that flows through an aggregate can lead to very different interactions between fractal-like particles compared to solid spherical particles (see e.g \cite{Stolzenbach1994,Li2001}), in particular such permeability effects may actually decrease the influence of hydrodynamic interactions between the aggregates compared to the case of completely solid particles \cite{Baebler2006}.  \\
A further limitation for the applicability of the model is the number of primary particles that can be computationally considered. The present computational capacities do not allow to apply this approach to large systems, e.g. models that are used to study aggregation, fragmentation and aggregate transport on spatial scales up to several hundred kilometers. For large systems a mean field approach is therefore much better suited. However, for many small systems and also for the principle study of the processes involved, this is not a severe limitation.\\
Furthermore, as the model was tested with a simple 3-d synthetic turbulent flow, it is an open question how representative the results are to draw general conclusions about the evolving steady state size distribution. In particular intermittency effects and clustering of particles on the inertial scale of a real turbulent flow may significantly affect aggregation and fragmentation probabilities. In order to achieve more general statements it will therefore be necessary to study the model and the resulting size distribution for various, more realistic flows, for example using DNS simulations of real turbulent flows. Nevertheless, the influence of different system parameters and fragmentation mechanisms has been tested and gives a detailed insight for this specific flow.\\

\subsection{Summary}
\label{sec:Summary}
In the present study we described in detail a coupled model for advection, aggregation and fragmentation of individual inertial particles with a fractal-like structure. We showed how typical properties of aggregation and fragmentation processes can be incorporated. In particular, we introduced an approximate way, using modified aggregate sizes and effective densities, to account for the fractal-like structure that is common for aggregates in many natural systems. The model represents an alternative approach to the mean field theory that is usually used to describe aggregation and fragmentation processes and was used to gain insights into principle behavior of fractal-like aggregates under different fragmentation mechanisms. The model was parameterized for the case of a suspension of marine aggregates in the ocean, but can in principle be used in a wide range of applications such as cohesive sediment dynamics, the flocculation of biological cells or solid-liquid separation systems in chemical engineering \cite{Kranenburg1994,Han2003,Baebler2008}. \\
We observed the development of a balance between aggregation and fragmentation, leading to a steady state. It was found that with increasing aggregate strength the mean aggregate size in steady state increases, whereas the relaxation time stays constant. With increasing fractal dimension the relaxation time towards steady state and the mean aggregate size in steady state increase. By contrast, an increase in particle volume fraction decreases the relaxation time due to higher collision probabilities and increases the steady state mean size. In general, increased aggregation rates or decreased fragmentation rates lead to an increased mean aggregate size in steady state. The relaxation time decreases for increasing aggregation rates, but does not change with decreasing fragmentation rates. \\
In the context of our model different types of fragment size distributions can easily be tested and compared with each other. We compared numerical results for three commonly used distributions of fragment sizes, large-scale fragmentation where fragments typically have similar sizes, erosion, where one fragment is typically very small and uniform fragmentation, where all fragment sizes appear with the same probability. Large-scale fragmentation and erosion were treated as two separate processes to study the influence of fragmentation at a distance $\zeta$ from the center of the aggregate on the size distributions in the steady state. In reality, aggregates will break with certain probabilities at a distance $\zeta$ from the center but there will not be two separate processes. Therefore, depending on the probability distribution for fragmentation at a certain distance, one can expect different combinations of the steady state size distributions found in this work. \\
One such combination, where fragmentation at all distances $\zeta$ appears with the same probability was given by the uniform fragmentation rule. Uniform fragmentation leads to a distribution with a broad plateau for small aggregate sizes and an exponential tail towards larger aggregate sizes. A distribution with two different regimes evolves for erosion-like fragmentation. Large-scale fragmentation leads to an exponential tail of the particle size distribution. Similar shapes of the size distribution of the aggregates for large-scale fragmentation have also been found in a number of theoretical and experimental studies \cite{Spicer1996,Mietta2008}, indicating that our model is able to reproduce the major features of such aggregation-fragmentation processes. Such an exponential tail has also been measured in field studies of marine aggregates \cite{Lunau2006}. This may indicate that large-scale fragmentation could be the primary mode of break-up for such aggregates, as has previously been discussed by e.g. Thomas et al. \cite{Thomas1999}. \\
In all cases the steady state particle size distribution follows a specific shape for each fragmentation rule. This indicates that the fragmentation process is most relevant for the shape of the distribution. The ratio of aggregation and fragmentation probabilities, mainly influenced by the aggregate strength, total particle volume fraction and fractal dimension, determines the mean aggregate size in steady state and the relaxation time. Out of these three parameters the fractal dimension has the strongest effect since it influences both aggregation and fragmentation probabilities.\\
The influence of large-scale fragmentation versus erosion for marine aggregates has recently been studied numerically and compared to experimental results in a work by Verney et al. \cite{Verney2010}. They used a mean-field based Smoluchowski equation approach and obtained results comparable to those of our model. Thus, as both model approaches lead to similar results the insight into fragmentation and fragment distributions provided by the perspective of our model can provide a useful addition to the understanding of aggregation and fragmentation processes. Additionally, in the particle-based model presented here particle inertia can be fully considered, while the correct incorporation of particle inertia into a mean field theory is still an unsolved problem. Hence, future model studies using this approach can lead to a better understanding of particle inertia effects in aggregation and fragmentation processes. Therefore, the model suggested here has the capability to be a powerful tool to investigate the validity of different approximative strategies in the formulation of a mean field theory. 

\section{Acknowledgments}
The authors thank A. Aldredge, T. T\'el, E. Villermaux, L.-P. Wang and M. Wilkinson for useful discussion and suggestions.


\begin{thebibliography}{10}
\bibitem{fractality}
Actually, one would have to relate the radius of a fractal-like aggregate to the
  fractal generator, i.e. the smallest structure with the same fractal
  properties instead of to the (spherical) primary particles. This would
  introduce additional shape factors and constants that are not present in the
  current form, but not change the qualitative results. For details on this,
  see e.g. Ref \cite{Logan}.
\end{thebibliography}
\end{document}